\documentclass{iau-JDSS}
\usepackage{graphicx}

\title[AGB stars from IPHAS]
{Galactic AGB stars from the IPHAS survey}

\author[N.J. Wright]
{N.J. Wright,$^1$ \\
M.J.~Barlow,$^2$ R.~Greimel,$^{3}$ J.E.~Drew$^{4}$ and M.~Matsuura$^{2}$}

\affiliation{
$^1$Harvard-Smithsonian Center for Astrophysics, 60 Garden Street, Cambridge, MA 02138\\
  $^2$University College London, Gower Street, London WC1E 6BT, U.K.\\
  $^3$Institut f\"ur Physik, Karl-Franzen Universit\"at Graz, Universit\"atsplatz 5, 8010 Graz, Austria\\
  $^4$University of Hertfordshire, College Lane, Hatfield, AL10 9AB, U.K.\\
}

\pubyear{2009}
\volume{Volume 15}
\pagerange{?-?}
\jname{The Galactic Plane, in depth and across the spectrum}
\editors{Janet Drew and Melvin Hoare, eds}
\begin{document}

\maketitle

\begin{abstract}

Asymptotic giant branch (AGB) stars are one of the final evolutionary stages of all intermediate mass stars and one of the major sources of enriched material returned to the interstellar medium, including all s-process elements and the majority of carbon. Quantitative knowledge of their chemistries and mass-loss rates is therefore vital for an understanding of galactic chemical evolution. The INT Photometric H$\alpha$ Survey (IPHAS, Drew et al. 2005) is imaging the entire northern Galactic plane using Sloan $r'$, $i'$, and narrow-band H$\alpha$ filters. The use of broad-band filters in the red makes this survey excellent at highlighting AGB stars whose cool photospheres emit predominantly in the red and near-IR. Additionally, one of the unique features of the IPHAS colour-colour diagram is that the main-sequence and giant branches are well separated at late spectral types, effectively allowing the AGB population across the entire northern Galactic plane to be identified and studied. Wright et al. (2008) presented a photometric analysis of the most extremely reddened sources in the IPHAS colour-colour diagram and confirmed that they were predominantly late-type AGB stars with high levels of circumstellar material that contributed to their reddening.

Follow-up spectroscopy on a number of optical and near-IR instruments has allowed this population to be studied in more detail. Wright et al. (2009) published a near-IR spectral library of AGB stars with a particular focus on very late-type sources. The spectral library includes spectra in all three near-IR bands as well as many variables objects and chemically evolved sources such as S-type and carbon stars. The library includes spectral classification sequences highlighting the various molecular features identified and discusses a number of rare features, for which the potential molecules responsible are discussed.

Finally we discuss a correlation between the IPHAS ($r' - $H$\alpha$) colour and the C/O abundance index (Keenan \& Boeshaar, 1980). Wright et al. (2009) found that the IPHAS ($r' - $H$\alpha$) colour could be used to estimate C/O ratios for S-type stars and therefore determine their state of chemical evolution in the transition from O-rich (C/O~$< 1$) to carbon rich (C/O~$> 1$) via the intermediate S-type phase (C/O~$\sim 1$). When combined with a near-IR colour the relationship has the potential to separate O-rich, S-type, and carbon stars across the Galactic plane based on their photometry alone. We discuss the potential benefits of such a sample on our understanding of AGB dredge-up mechanisms, galactic chemical evolution, and the structure and metallicity of the Galactic disk.


\end{abstract}


\begin{thebibliography}{}

\bibitem[Drew et al. (2005)]{drew05}
{Drew, J.~E., Greimel, R., Irwin, M.~J., et al.,} 2005,
\textit{MNRAS}, 362, 753

\bibitem[Keenan \& Boeshaar (1980)]{keen80}
{Keenan, P.~C. \& Boeshaar, P.~C.} 1980
\textit{ApJS}, 43, 379

\bibitem[Wright et al. (2008)]{wrig08}
{Wright, N.~J., Greimel, R., Barlow, M.~J., et al.,} 2008,
\textit{MNRAS}, 390, 929

\bibitem[Wright et al. (2009)]{wrig09}
{Wright, N.~J., Barlow, M.~J., Greimel, R., et al.,} 2009,
\textit{MNRAS}, 400, 1413
     
     
     
     
     
     
     
     
     
     

\end{thebibliography}
\end{document}